\documentclass{article}
\usepackage[T1]{fontenc} % add special characters (e.g., umlaute)
\usepackage[utf8]{inputenc} % set utf-8 as default input encoding
\usepackage{ismir,amsmath,cite}
\usepackage{graphicx}
\usepackage{svg}
\usepackage{color}
\usepackage{import}
\usepackage{amsfonts}
\usepackage{lineno}
\usepackage{soul,color}
\usepackage{multirow}
\usepackage{makecell} %hline tables thickness
\usepackage{subcaption}
\usepackage{caption}
\usepackage[export]{adjustbox}
\usepackage{booktabs}
\usepackage{float}
\usepackage{amsmath}
\usepackage[hyphens]{url} 
\usepackage[bookmarks=false,hidelinks]{hyperref}

\title{Timbre Classification of Musical Instruments with a Deep Learning Multi-Head Attention-Based Model}

\oneauthor
{Carlos Hernandez-Olivan, Jose R. Beltran}
{Universidad de Zaragoza \\
{\tt \{carloshero, jrbelbla\}@unizar.es }}

\begin{document}

\maketitle
\begin{abstract}
The aim of this work is to define a model based on deep learning that is able to identify different instrument timbres with as few parameters as possible. For this purpose, we have worked with classical orchestral instruments played with different dynamics, which are part of a few instrument families and which play notes in the same pitch range. It has been possible to assess the ability to classify instruments by timbre even if the instruments are playing the same note with the same intensity.
The network employed uses a multi-head attention mechanism, with 8 heads and a dense network at the output taking as input the log-mel magnitude spectrograms of the sound samples. This network allows the identification of 20 instrument classes of the classical orchestra, achieving an overall F$_1$ value of 0.62. An analysis of the weights of the attention layer has been performed and the confusion matrix of the model is presented, allowing us to assess the ability of the proposed architecture to distinguish timbre and to establish the aspects on which future work should focus.

\end{abstract}

%----------1.INTRO-----------
\section{Introduction}\label{sec:introduction}
\begin{figure*}
    \centering
    \includegraphics[width=0.9\textwidth, height=5.5cm]{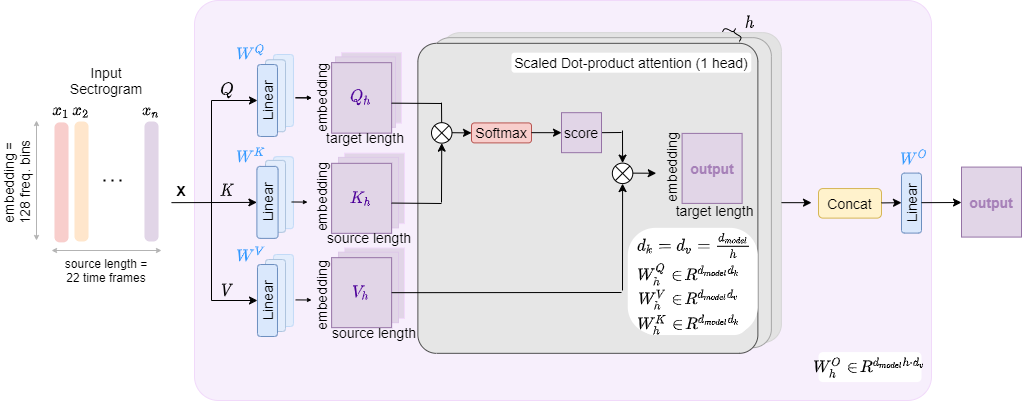}
    \caption{Multi-Head Attention block diagram (biases are ignored).}
    \label{fig:attention}
\end{figure*}
%The estimation of the fundamental frequency ($f_0$), pitch estimation or pitch tracking has been studied for decades and it is one of the main topics of research in the Music Information Retrieval (MIR). Pitch estimation include monophonic and polyphonic audio signals, and it has been used recently in a wide-variety of tasks such as speech recognition systems \cite{ghahremani2014pitch}, to annotate multi-track datasets \cite{medleydb}. We will use pitch and fundamental frequency interchangeably in this paper although pitch does not always correspond to the physical property of the fundamental frequency.

%Pitch estimation has been studied for decades. Traditional methods used a function to 

Timbre has been studied for decades from the psychoacoustics and music psychology point of view, and also from the perspective of signal processing. For extracting timbre from audio signals, it is necessary to extract quantitative descriptors, which are referred as audio \textit{features} in the Music Information Retrieval (MIR) tasks \cite{timbre_book}. In MIR, extracting audio features is an important part of most of research fields \cite{casey2008}, \cite{levy2009}. The importance of the study of timbre relies on the depency of other sound properties or MIR research areas on it, such as the fundamental frequency \cite{marozeau2003dependency}, the music structure \cite{mcadams1999perspectives} or the instrument recognition tasks.

Firsts studies which tried to model timbre were based on timbre spaces. The term \textit{timbre space} was born in the 1970s with the works of Grey \cite{grey1977multidimensional}, Grey and Moorer \cite{grey1977perceptual} and later in the 2000s, with Iverson and Krumhansl's work \cite{iverson1993isolating}. This approaches recorded and modified instrument tones in order to build a \textit{timbre space} which evaluated the perceptual relationships between music instrument tones by building an interpetable cluster in terms of spectral, temporal and spectrotemporal properties of sound events. In 2006, McAdams et al. \cite{mcadams2006meta} described the correlation of continuous perceptual dimensions with acoustic parameters such as spectral, temporal and spectrotemporal properties of sound events. They define timbre space as a model that predicts other perceptual results such as auditory stream formation.

Recognizing instruments with deep neural networks, is an active area of research in the field of Music Information Retrieval (MIR). Previous deep learning models that have been proposed in the recent years to recognize instruments by their timbre used Convolutional Neural Networks (CNNs) as the preferred architecture for modeling timbre. CNNs are a common neural network architecture which is used in lots of MIR tasks due to their 2-dimensional filters that allow to extract features in the time-frequency domain. In timbre modeling, CNNs allows to interpret time and frequency invariances with a small number of parameters, therefore, they are very useful in modeling timbre. Previous studies that modelled timbre with CNNs use different filter sizes such as small-rectangular filters \cite{ChoiFS16}\cite{HanKL17} or high dimensional filters \cite{DielemanS14}\cite{LeePLN09}. The problem of those models is that either the models are too small to learn timbre, or they are too large so the model overfits. To solve that, more recent works use different filter sizes in the first layer of the CNN \cite{pons_timbre_cnn}. This approach allows to model relevant time-frequency contexts with a small number of parameters and preventing the model to overfit.

In our work, we show the possibility to model timbre with end-to-end deep learning using the log-mel magnitude spectrogram as the input of our models. In our case, we do not need to build manually pre-processing steps to obtain sound descriptors nor take assumptions over those descriptors in order to model timbre. In addition, we classify instruments of the classical music orchestra, which means that some of those instruments such as the violin and the viola have similar timbres and the notes they played belongs almost to the same pitch. Our work is structured as follows: in section \ref{sec:introduction} (current section) we introduce the concept of timbre, in section \ref{sec:related_work} we describe the state-of-the-art, in section \ref{sec:proposed_method} we briefly describe the self-attention mechanism and we present our model, in section \ref{sec:evaluation} we explain the experiments that have been done and in section \ref{sec:conclusions} we give the conclusions and the future work.

We show further results of our work in a website\footnote{\url{https://carlosholivan.github.io/publications/2021-timbre/2021-timbre.html}} and made our code publicly available\footnote{\url{https://github.com/carlosholivan/Timbre-Classification-MultiHeadAttention}}.

%----------2.RELATED WORK-----------
\section{Related Work}\label{sec:related_work}

Studies for timbre identification can be divided in two big groups according to the techniques used for timbre analysis: traditional methods and deep learning methods. Traditional methods model timbre by finding the adequate sound descriptors whereas deep learning methods are usually end-to-end methods in which timbre is learned by a deep neural network. We briefly describe these methods below.

\subsection{Traditional Methods}
As we mention in Section \ref{sec:introduction}, timbre has been studied for decades. Traditional methods for timbre modeling started in the 1970s. These approaches recorded and modified instrument tones in order to build a \textit{timbre space} which evaluated the perceptual relationships between music instrument tones by building an interpetable cluster in terms of spectral, temporal and spectrotemporal properties of sound events. Other works studied this problem by recording musical instrument tones and combining sound descriptors in different dimensions to form a combined timbre space \cite{lakatos2000common}, and others built the timbre space by making a cluster tree of a set of 72 descriptors \cite{mcadams2006meta}. The descriptors taken to build the timbre space determined the nature of the different dimensions of it \cite{siedenburg2019timbre}. 
More recent methods use unsupervised algorithms such as $k$NN \cite{kaminskyj2005automatic} and counter propagation neural networks applied to the MFCCs \cite{BhalkeRB16} in order to be able to classify more instruments and to distinguish instruments from the same instrument families. There have been also proposed approaches to classify intruments by their playing techniques \cite{AgostiniLP03}. The limitation of these methods is that it is necessary to compute and select a high number of audio features in order to find the ones that model timbre, which in a deep learning end-to-end model is learned by the neural network.

\subsection{Deep Learning Methods}
In the recent years, with the growth of deep learning models, studies are focusing on identifying the most predominant instrument in mixtures (signals where multiple instruments are present). Although instrument recognition in monophonic recordings of isolated instruments have been studied for years \cite{WieczorkowskaC03}\cite{LostanlenAL18}, there is still a challenge when it comes to identify instruments with similar timbres with end-to-end methods.
Latest models focus on recognizing the instruments which are present in polyphonic music signals. As we mention in Section \ref{sec:introduction}, CNNs have been the predominant neural network architecture for this task. Han et al. \cite{HanKL17} used CNNs for instrument recognition on the IRMAS dataset \cite{BoschJFH12} and Li et al. \cite{LiQW15} applied CNNs to raw audio to identify instruments of the MedleyDB dataset \cite{BittnerSTMCB14}. Multi-task deep learning has been also used in these tasks in order to detect instruments and pitch. Hung et al. \cite{HungY18} proposed a multi-task deep learning approach that outperforms previous approaches which used MusicNet dataset \cite{ThickstunHK17} for pitch conditioning, which is demonstrated to improve instrument recognition results.  

Other techniques try to model timbre by using magnitude spectrograms as the input of the models \cite{solanki2019music} \cite{pons_timbre_cnn}, and augmentation techniques in order to increase the number of training samples \cite{kratimenos2021augmentation}.
Pons et al. \cite{pons_timbre_cnn} describe the design of a  learning model that learn timbre: the model should be pitch, loudness, duration and spatial position invariant. Timbre modeling is not only useful to recognize istruments but it is also important for other MIR tasks such as sound synthesis or Automatic Music Transcription (AMT).
New approaches for multi-label instrument
recognition uses the attention mechanism in order to predict the presence of instruments in weakly labeled datasets \cite{GururaniSL19}.

We propose a model which is able to classify monophonic sound samples by their timbre. We use a dataset (see Section \ref{sec:data}) with 20 instrument classes that have been recorded with a wide-variety of dynamics, techniques such as pizzicato and vibrato, and notes. We demonstrate how the model is pitch invariant which means that it understands timbre separately than pitch. 

%----------3.PROPOSED METHOD-----------
\section{Proposed Method}\label{sec:proposed_method}
%\externaldocument{2-related_work.tex} %include fig which is in file 2

In this section, we first give a short description of the self-attention mechanism and then we describe our model. In Fig. \ref{fig:attention} we show the self-attention block diagram scheme.

\subsection{Self-Attention} \label{sec:vae}
The self-attention mechanism along with the Transformer model was introduced by Vaswani et al. in 2017 \cite{attention}. This model has become one of the most important models in the recent years in Natural Language Processing (NLP) applications. 
We consider a set $n$ of inputs $\mathrm{x}=\left\lbrace\mathrm{x}_1,...,\mathrm{x}_n\right\rbrace$. 
The attention function creates set of vectors that are called keys $k$, queries $q$ and values $v$, for each input. These vectors are packed into matrices $K, Q, V$ respectively. The vectors $k, q, v$ are obtained by multiplying the embedding of each input by three matrices $W^K, W^Q, W^V$ which are associated to these vectors and that are learnt during the training process.
Then, a score $\alpha$ is computed to let the model focus on the relevant positions of each input in the sequence. After the scores for each input are computed, they are divided by $\sqrt{d_k}$ and passed through a softmax activation layer. After that, the obtained softmax score is multiplied by the values $v$ in order to discard irrelevant inputs and retain the important ones. To finish with, the resulting values are weighted and summed. The key, queries and values from which attention is computed are vectors of dimension $Q \in \mathbb{R}^{d_{model}\times d_k}$, $V \in \mathbb{R}^{d_{model}\times d_v}$ and $K \in \mathbb{R}^{d_{model}\times d_k}$ with $d_{model}$ the model's size, $d_k = d_v = d_{model} / h$ and $h$ the number of heads.
In Eq. \ref{eq:score} we show the general expression of the so called scaled dot-product attention \cite{attention}.

\begin{equation}
    \mathrm{Attention}(Q,K,V) = \mathrm{softmax}\left( \frac{QK^\top}{\sqrt{d_k}}\right)V
    \label{eq:score}
\end{equation}

Multi-Head attention allows to perform attention in parallel and attend information from different subspaces at different positions. In Eq. \ref{eq:attention} we show the general expression of Multi-Head Attention \cite{attention}.

\begin{equation}
    \mathrm{MultiHead}(Q,K,V) = \mathrm{Concat}(h_1,...,h_n)W^0
    \label{eq:attention}
\end{equation}

where $h_i$ is the i$^{t}$ head attention: $h_i = \mathrm{Attention}(QW^Q_i, KW^K_i, VW^V_i)$. 

\subsection{Pre-Processing: Inputs}
We use mel magnitude spectrograms as the inputs of our model. To compute the mel spectrograms, we use a sample rate of 22050Hz, 128 frequency bins in the range between 32.7Hz and 8000Hz, a hop size of 512 samples and an overlap of 50\%.
Then, in order to obtain inputs of the same size, inspired by \cite{gmmvae} we only take the first 500ms of each sound sample, which corresponds to 22 time frames at 22050Hz.
The sound samples of the dataset we use (see Section \ref{sec:data}) start with silence in the first miliseconds of the recordings. We remove that silence in order to take the 500ms where the instrument starts playing. To do that, after computing the log-mel-magnitude-spectrograms, we select an energy threshold of 0.1, so the time frame where the sound sample starts is the first time frame where we found the first energy value which is higher than 0.1. Once we find this time frame, we take the following 22 time frames of the magnitude-spectrogram.
Then, we normalize the magnitude spectrogram to zero mean and unit variance in the frequency dimension.
We use \textit{librosa}\footnote{\url{https://github.com/librosa/librosa}, accessed May 2021} library \cite{mcfee2015librosa} to compute the inputs of our model.

\subsection{Model Architecture}

\begin{table*}[h!]
\small
    \centering
    \begin{tabular}{>{\centering\arraybackslash}p{3cm} >{\centering\arraybackslash}p{1.6cm} >{\centering\arraybackslash}p{1.25cm} c c c c}
    \toprule
        Model & total params. & num. params. & Layer & Parameters & Input & Output\\
        \midrule
        \multirow{6}{2.5cm}{Freq.\ Attention} & \multirow{2}{2.5cm}{122,388} & 66,048 & att1 & $h$:1 & [$b$, 1, 128, 22] & [22, $b$, 128]\\
        & & 56,340 & fc & in:128x22, out:20 classes & [$b$, 128x22] & [$b$, 20] \\
        \cline{2-7}
        & \multirow{2}{2.5cm}{122,388} & 66,048 & att1 & $h$:8 & [$b$, 1, 128, 22] & [22, $b$, 128]\\
        & & 56,340 & fc & in:128x22, out:20 classes & [$b$, 128x22] & [$b$, 20]\\
        \cline{2-7}
        & \multirow{2}{2.5cm}{122,388} & 66,048 & att1 & $h$:16 & [$b$, 1, 128, 22] & [22, $b$, 128]\\
        & & 56,340 & fc & in:128x22, out:20 classes & [$b$, 128x22] & [$b$, 20] \\
        \midrule
        \multirow{2}{2.5cm}{Freq. FC} & \multirow{2}{2.5cm}{72,852} & 16,512 & fc1 & in:128, out:128 & [$b$, 1, 128, 22] & [$b$, 128, 22]\\
        & & 56,340 & fc2 & in:128x22, out:20 classes & [$b$, 128x22] & [$b$, 20] \\
         \bottomrule
         %\cmidrule{3-4}
    \end{tabular}
    \caption{Model's summaries of the architectures proposed in our work. Parameter $b$ corresponds to the batch size.}
    \label{tab:models}
\end{table*}

\begin{figure}[h]
    \centering
    \includegraphics[width=0.45\textwidth]{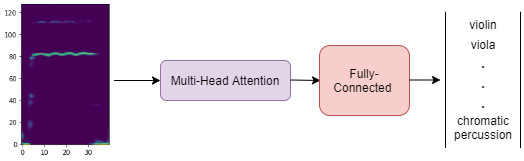}
    \caption{Freq.\ Attention model's architecture used in this work}
    \label{fig:model}
\end{figure}

In this work we propose two models which we call Freq.\ FC and Freq.\ Attention models. The Freq.\ FC model is composed by one Fully Connected (FC) layer followed by a ReLU activation function that works in the frequency axis and a second fully connected layer which takes the flattened output of the first layer. The second model we propose is composed by a Multi-head Attention layer \cite{attention} and a fully connected layer. The model architecture is shown in Fig.\ \ref{fig:model}.
The attention mechanism is applied by setting the embedding dimension equal to the frequency bins of the magnitude spectrogram (128 bins) and the length of the sequence equal to the number of timbre frames (22 frames) in order to let the model learn frequency spectral features, so it is pitch invariant. This translated to the NLP area can be understood as if our time frames (columns in the magnitude spectrogram) were words with their embedding dimension being the frequency bins of the spectrogram, and therefore all the time frames would be a "sentence" composed by 22 words (22 time frames).
Therefore, in our model, because we set $d_{model} = 128$ (the embedding dimension), the key, queries and values are vectors of dimension $Q \in \mathbb{R}^{128\times d_k}$, $V \in \mathbb{R}^{128\times d_v}$ and $K \in \mathbb{R}^{128\times d_k}$ with $d_k = d_v = 128 / h$ and $h$ the number of heads.

We use Multi-Head Attention so each head can focus on different spectral features. We detail how the models behaves when varying the number of attention heads in Section \ref{sec:evaluation}, and we compare the performance of the Freq.\ FC model against the Freq.\ Attention model.
After the attention layer, we flatten the output vector of its output so we get a vector with the information that the attention layer has extracted from the frequency dimension and the time frames of the magnitude spectrum. We pass this vector to a fully connected layer with the number of outputs equal to the instrument classes in order to get the scores of each instrument class. The parameters of each model we trained are described in Table \ref{tab:models}.

%----------4.EVALUATION AND RESULTS-----------
\section{Evaluation and Results}\label{sec:evaluation}

\begin{figure}[h]
    %------violin---------
    \begin{subfigure}[t]{0.2\linewidth}
        \centering
         \includegraphics[scale=0.45]{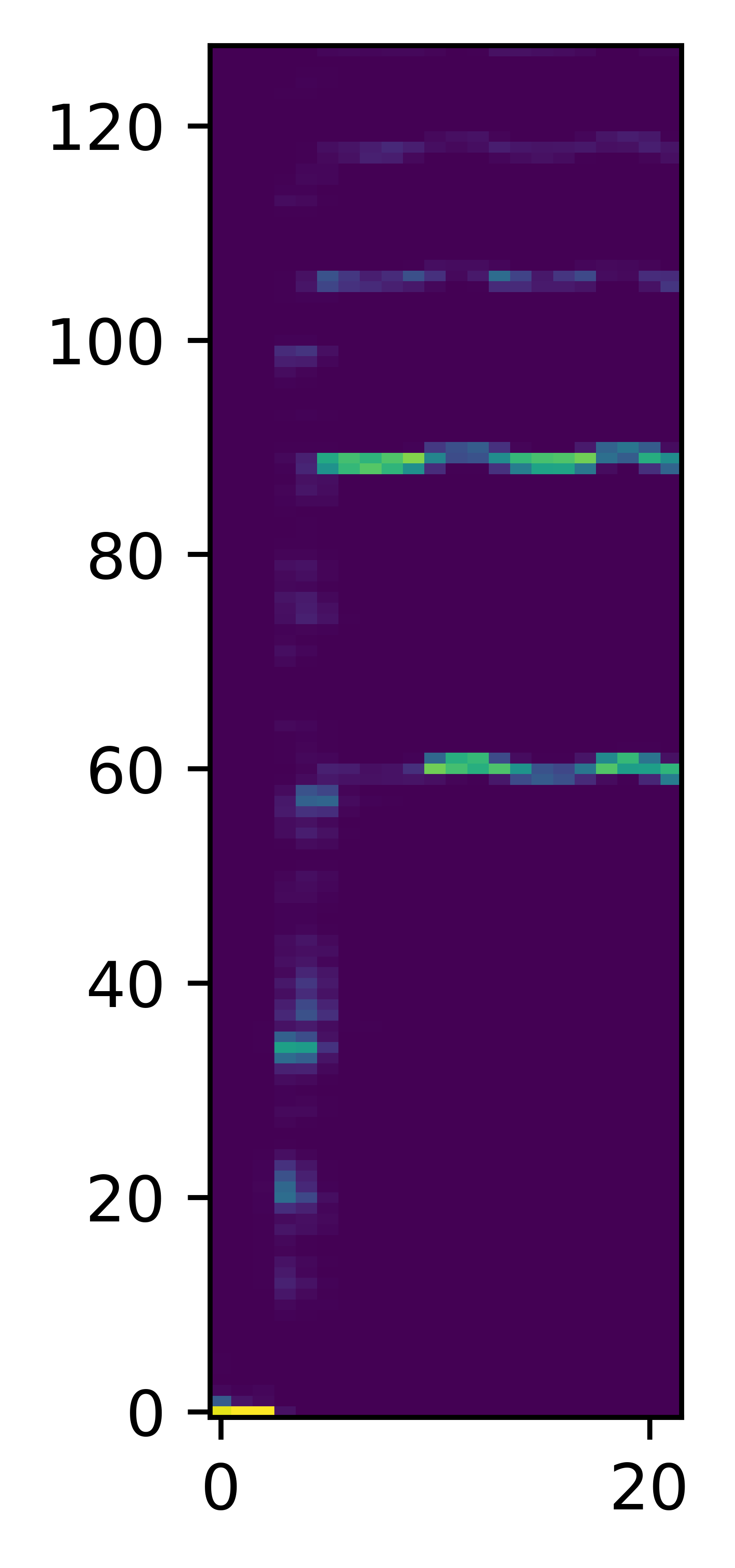}
         \label{fig:violin}
     \end{subfigure}
     \begin{subfigure}[t]{0.5\linewidth}
         \centering
         \includegraphics[scale=0.45]{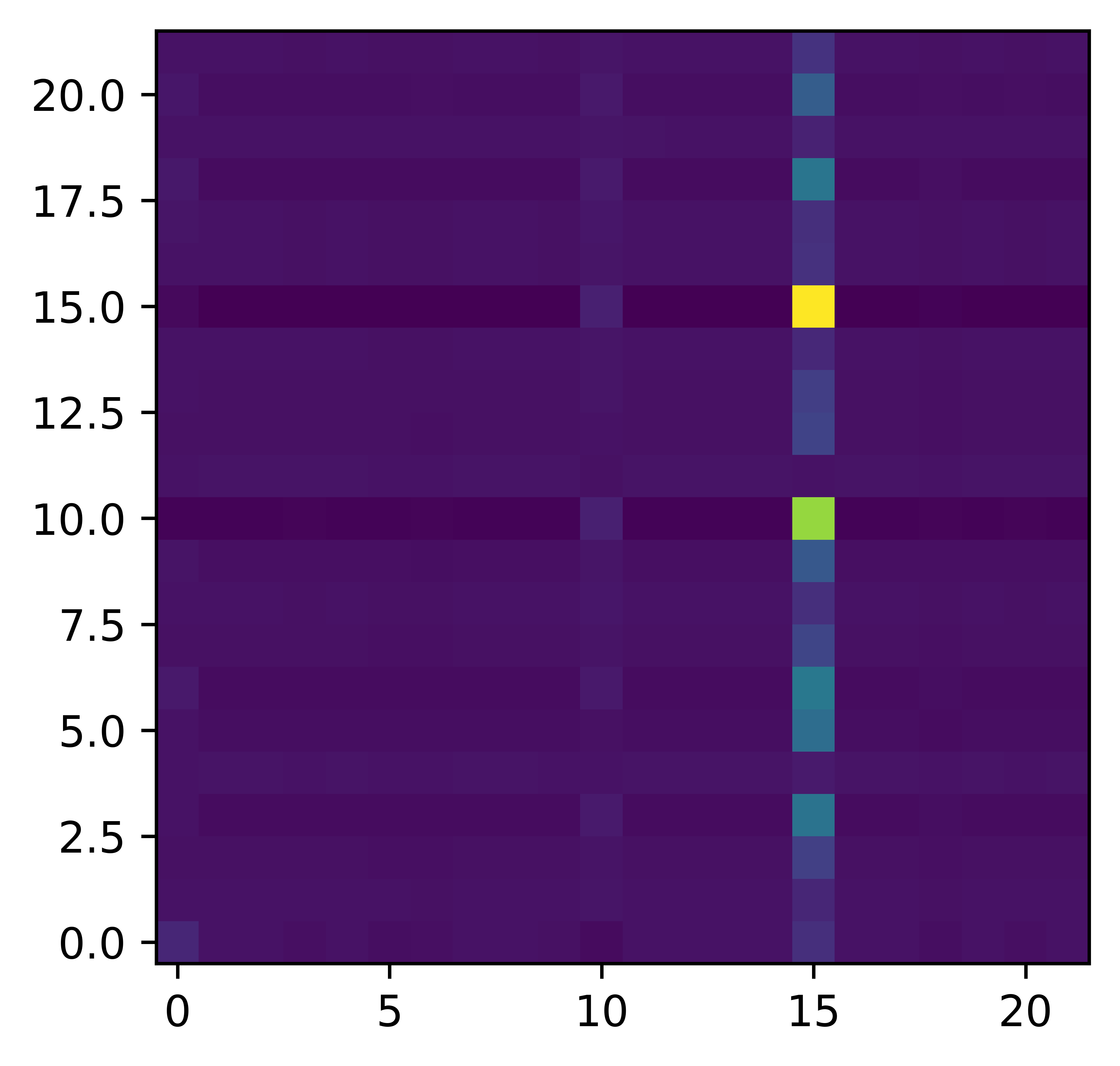}
         \label{fig:violin_weigths}
     \end{subfigure}
     \begin{subfigure}[t]{0.2\linewidth}
         \centering
         \includegraphics[scale=0.45]{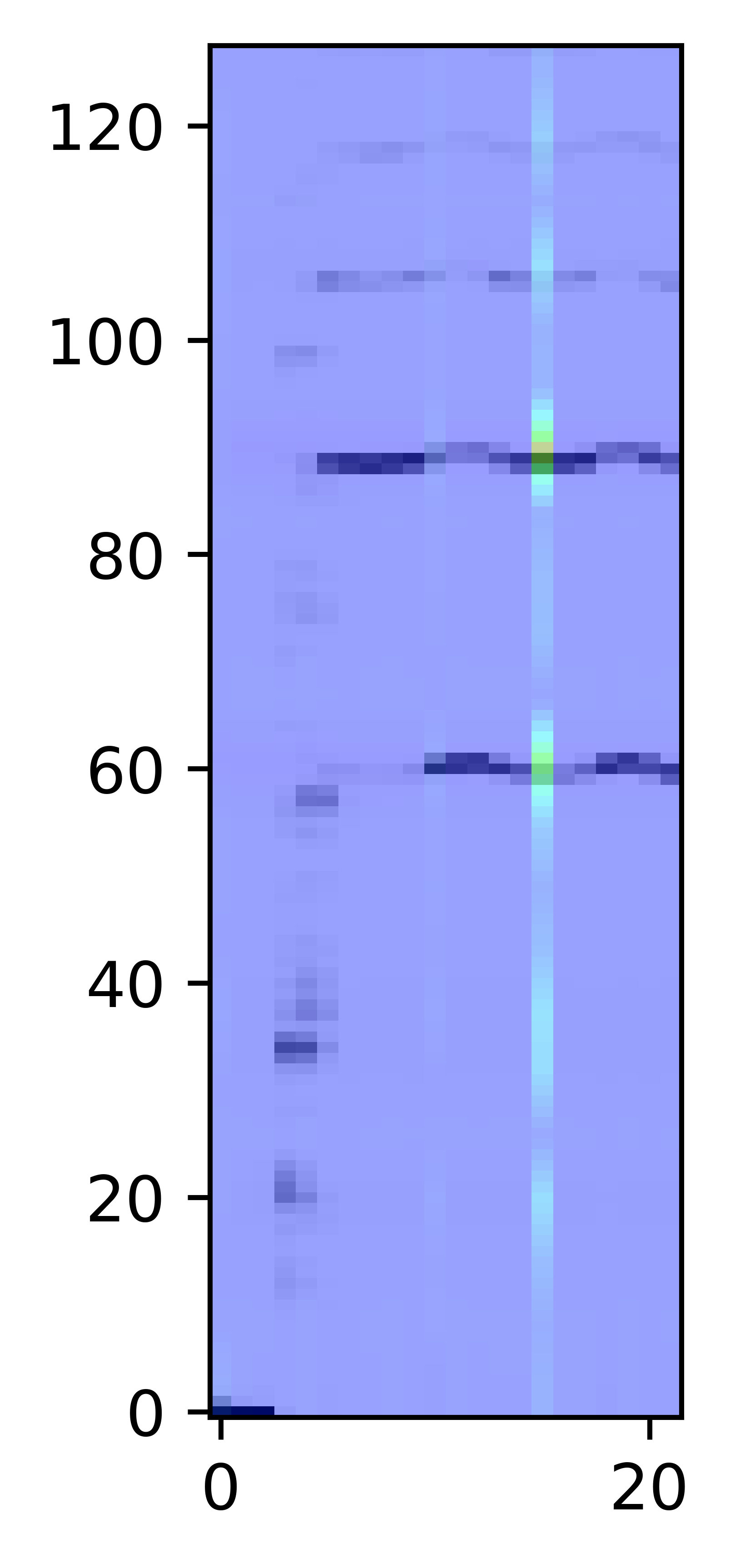}
         \label{fig:violin_map}
     \end{subfigure}
     \newline
     %------viola---------
     \begin{subfigure}[t]{0.2\linewidth}
         \centering
         \includegraphics[scale=0.45]{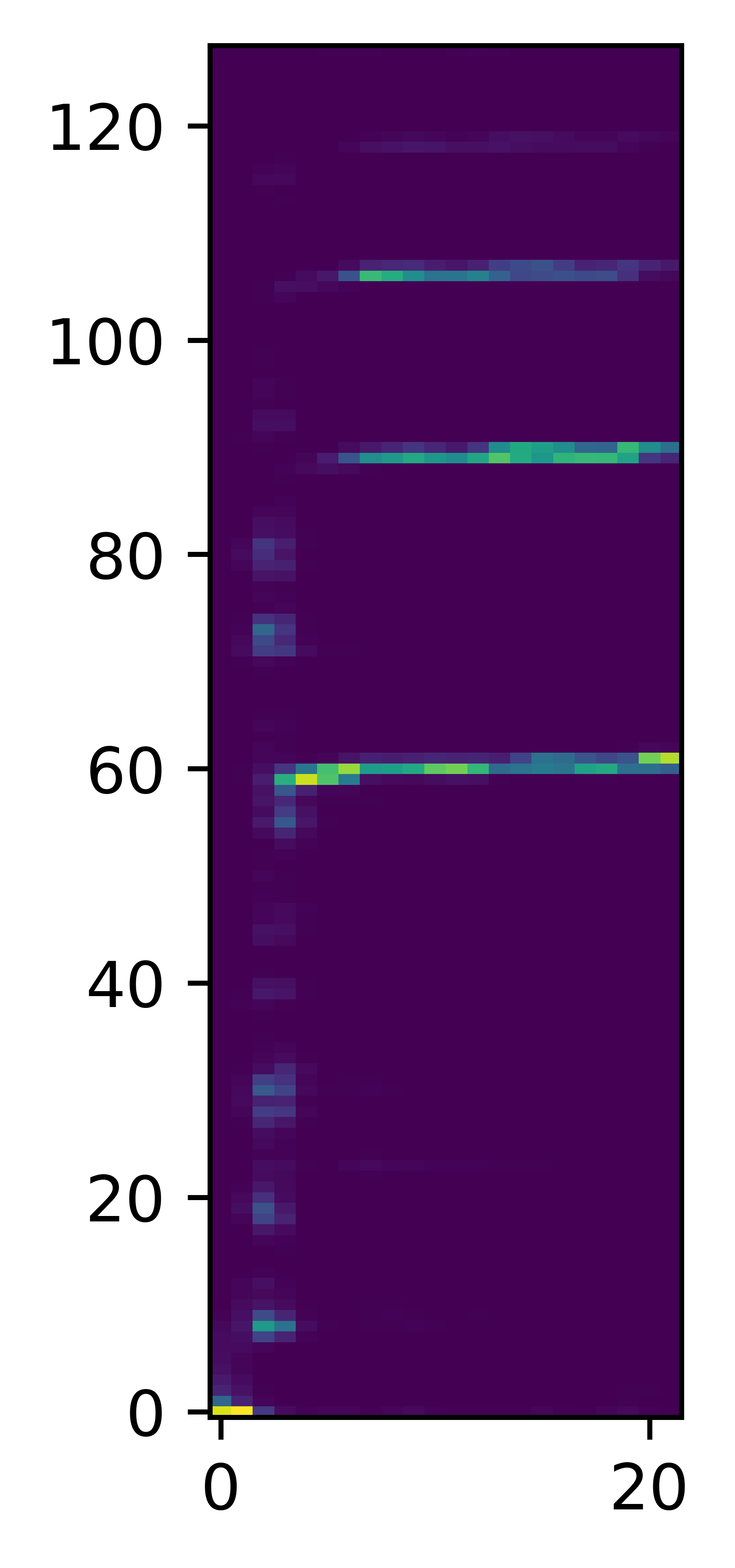}
         \label{fig:viola}
     \end{subfigure}
     \begin{subfigure}[t]{0.5\linewidth}
         \centering
         \includegraphics[scale=0.45]{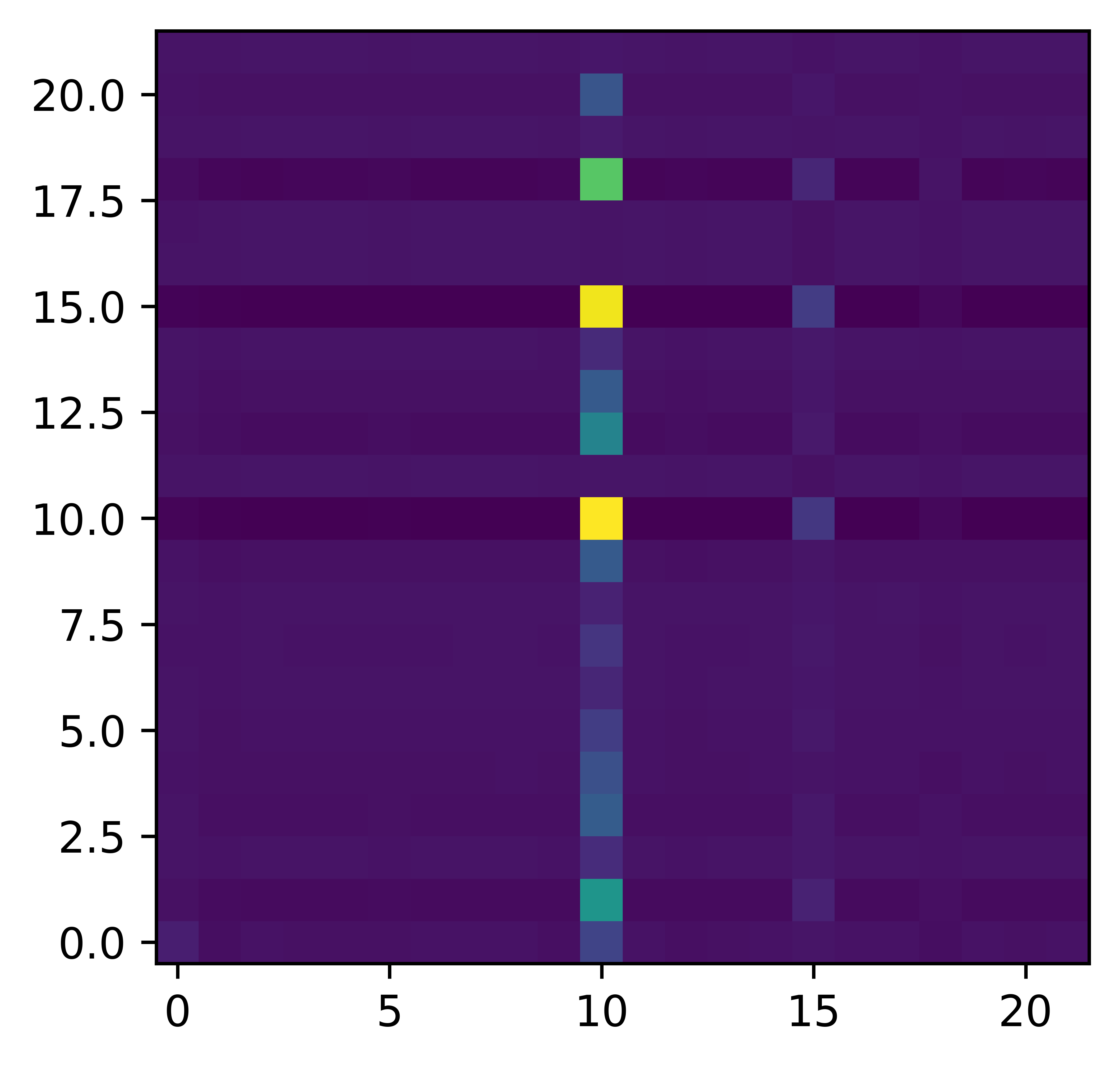}
         \label{fig:viola_weigths}
    \end{subfigure}
    \begin{subfigure}[t]{0.2\linewidth}
         \centering
         \includegraphics[scale=0.45]{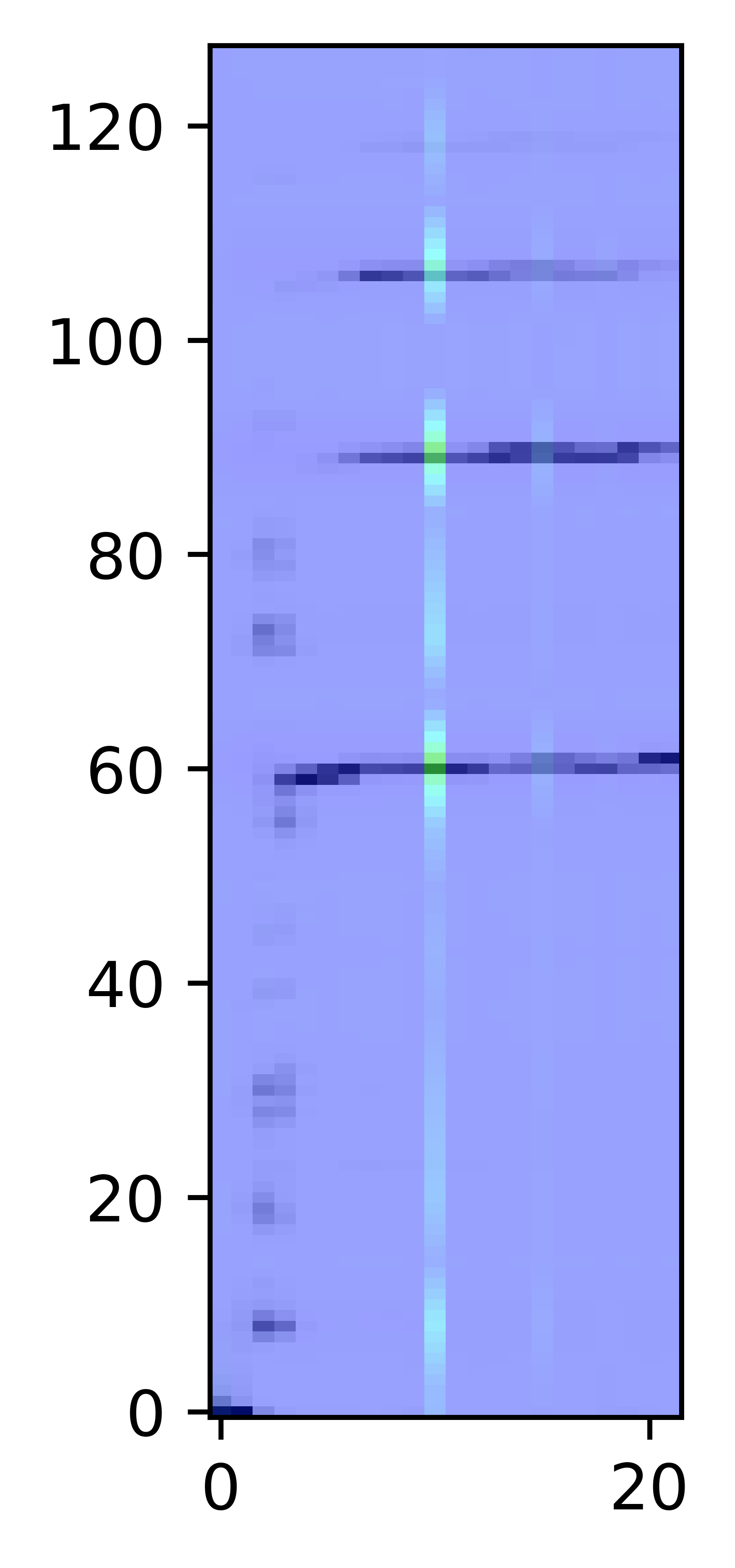}
         \label{fig:viola_map}
     \end{subfigure}
\caption{Violin sample from the London Philarmonic Orchestra (up left) with the average attention weights of all attention heads (up middle), and the attention activations (up rigth). A viola sample is showed down.}
\label{fig:att_weights}
\end{figure}

\subsection{Dataset} \label{sec:data}
For our experiments, we use the public available London Philarmonic Orchestra Dataset\footnote{\url{https://philharmonia.co.uk/resources/sound-samples/}, accessed on May 2021}. The dataset contains around 13700 monophonic sound samples of 58 different instruments. The dataset is divided in the following instrument families: woodwind, brass, percussion and strings. Woodwind instruments are bass clarinet, clarinet, bassoon, contrabassoon, flute, oboe and saxophone, string instruments are violin, viola, cello and double-bass, guitar, mandolin and banjo, brass instruments are french horn, english horn, trombone, trumpet and tuba and 39 percussion instruments such as agogo bells, bass drum and snare drum, which we group in a single class called chromatic percussion.
The sound samples are recorded with different techniques such as tremolo, pizzicato for bowed string instruments and a wide-range of dynamics such as piano, mezzo-piano, mezzo-forte and forte.
In our experiments, we use 13681 sound samples which we divide in 70\% train (11630 samples), validation 10\% (694) and 20\% test (1357 samples). The number of instruments of each class are shown in Table \ref{tab:results}.
The dataset annotations of pitch, instrument, dynamic and playing technique are written in the sound sample's filenames separated by underscores. For our study, we do take only into account the instrument label.

\begin{figure*}[h!]
%\captionsetup{justification=centering}
     \begin{subfigure}[t]{0.5\textwidth}
         \centering
         \includegraphics[scale=0.4]{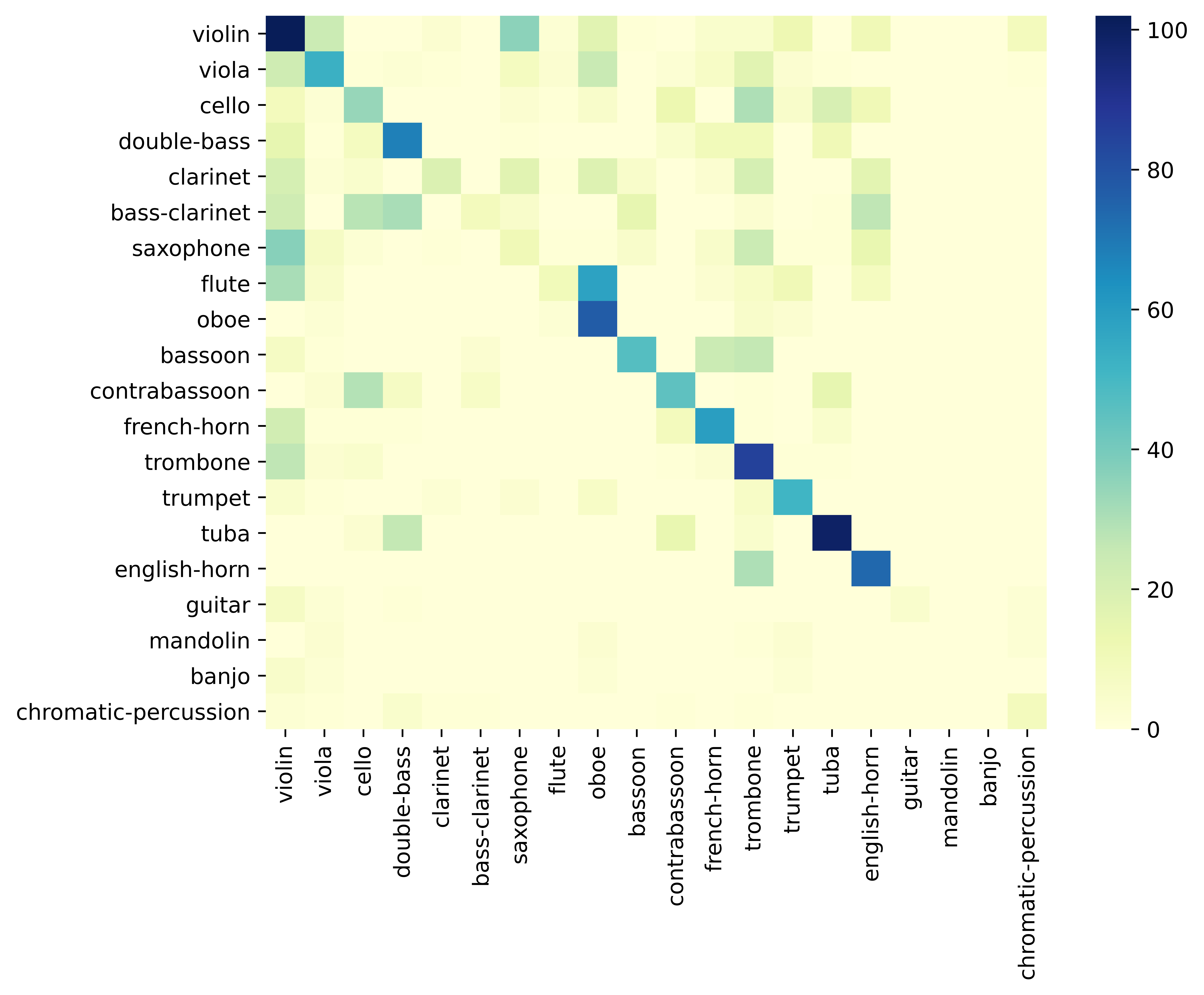}
         \caption{\centering}
         \label{fig:confusion_matrix_fc}
     \end{subfigure}
     \begin{subfigure}[t]{0.5\textwidth}
         \centering
         \includegraphics[scale=0.4]{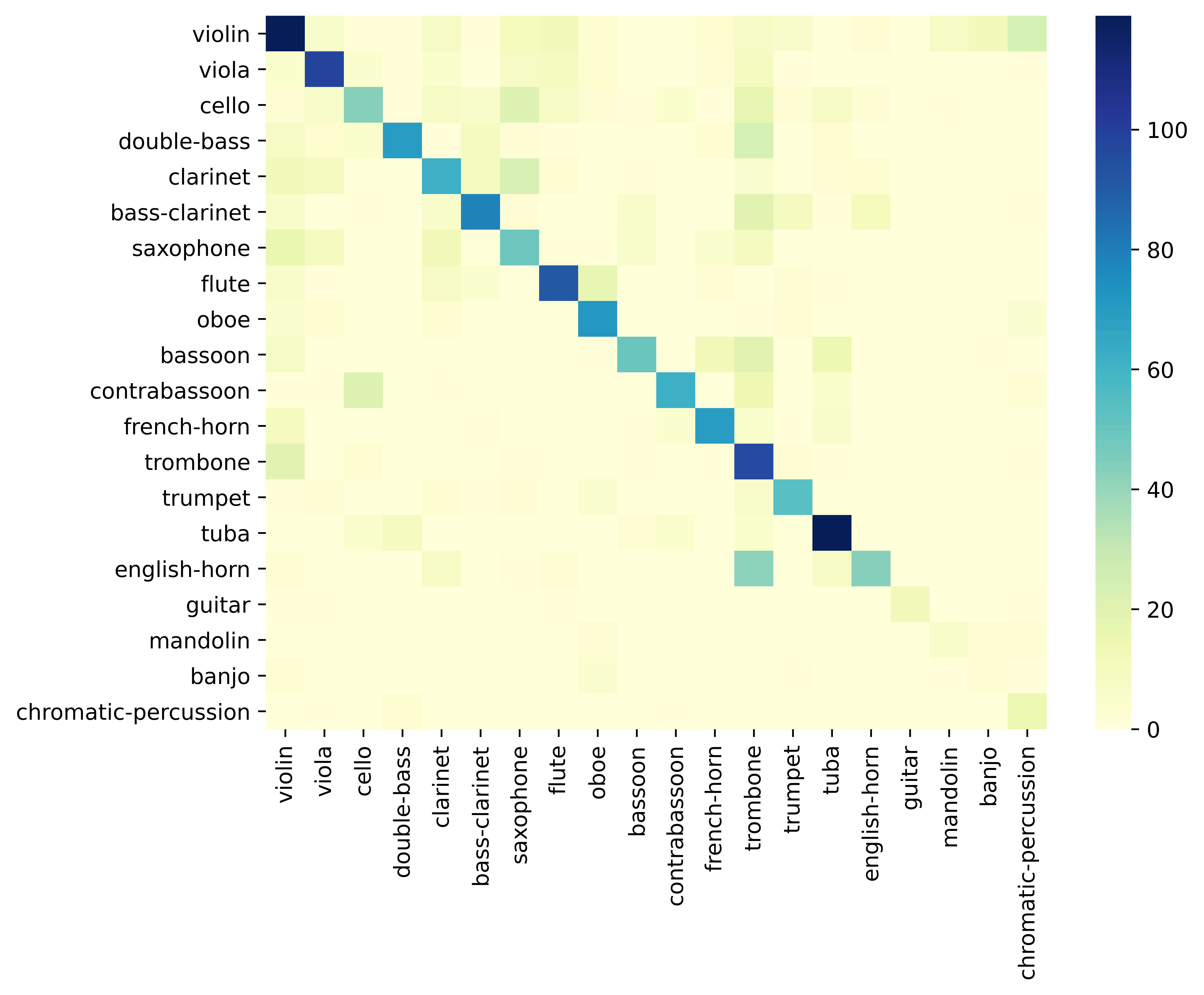}
         \caption{\centering}
         \label{fig:confusion_matrix_att}
     \end{subfigure}
\caption{Confusion matrices for our test set with the Freq. FC model (left) and the Freq. Attention model with $h$=8 (right).}
\label{fig:confusion_matrices}
\end{figure*}

\subsection{Training}
We train our model with 11630 sound samples. We use the Cross Entropy Loss $L$ for multi-classification problem which expression is shown in Eq. \ref{eq:cross_entropy}. 

\begin{equation}
    \label{eq:cross_entropy}
    L = -y \cdot log(\hat{y}) - (1-y) \cdot log(1-\hat{y})
\end{equation}
where $\hat{y}$ is the model's prediction and $y$ is the class label.

We use a batch size of 16, a learning rate of $10^{-5}$, a weight decay of $10^{-5}$ and the Adam optimizer \cite{kingma2015adam}.
Using weight decay as a regularization technique prevents the model to focus only on spectral features where the energy is higher, namely, the loudness. Therefore this allows the model to be loudness invariant \cite{pons_timbre_cnn}.
We train our model with different number of attention heads $h$ in order to show how this hyperparameter improves the overall performance of the model, and to show how many parameters does the model need to learn a well representation of timbre.

\subsection{Metrics}
We evaluate the results with Precision (P), Recall (R) and F-score (F$_1$) metrics. Because of the London Philarmonic Orchestra is not a commonly used dataset in MIR, the number of sound samples per instrument is not well balanced. The number of instruments and our training, validation and test sets are shown in Table \ref{tab:results}. Therefore, in order to give importance to the prediction itself based on the proportion of each instrument in the dataset, we give the total metrics of the classification by computing the weighted average of all the instruments.

\subsection{Ablation Study}
We train our model with different number of attention heads $h$. In Table \ref{tab:model_results} we show the results for every experiment done in this work. The values of P, R and F$_1$ are the average values of all the 20 classes of instruments. 
Analyzing the results, we see that the Freq.\ Attention model performs much better than the Freq.\ FC model. For the Freq.\ Attention model, the number of heads in the attention layer affects the results significantly. When $h$=1 (one attention head), the averaged F$_1$ value for the 20 instrument classes is 0.33. If we increment $h$ the performance of the model increases but when $h$=16 the model starts overfitting, which is something that is reasonable \cite{MichelLN19}. We found that the optimum value of $h$ is 8, for which the loss reaches its minimum value and the F$_1$ is the highest among our experiments with a value of 0.62.

\begin{table}[h]
\small
    \centering
    \begin{tabular}{>{\centering\arraybackslash}p{3.3cm} c |c c c}
    \toprule
        & & \multicolumn{3}{c}{Metrics}\\
        Model & Loss & P & R & F$_1$\\ 
        \midrule
        \midrule
        Freq.\ Attention ($h$=1) & 1.70 & 0.43 & 0.35 & 0.33 \\
        Freq.\ Attention ($h$=8) & 1.56 & 0.66 & 0.62 & 0.62 \\
        Freq.\ Attention ($h$=16) & 1.66 & 0.54 & 0.48 & 0.48 \\
        \midrule
        Freq. FC & 1.94 & 0.45 & 0.42 & 0.39 \\
         \bottomrule
         %\cmidrule{3-4}
    \end{tabular}
    \caption{Weighted average metrics for the 20 instrument classes and for each model.}
    \label{tab:model_results}
\end{table}

\begin{table*}[t]
\small
    \centering
    \begin{tabular}{c c |c c c |c c c |c c c}
    \toprule
        & & \multicolumn{3}{c|}{Dataset} & \multicolumn{3}{c|}{\makecell{Metrics\\ Freq. Attention ($h$=8)}} & \multicolumn{3}{c}{\makecell{Metrics\\ Freq. FC}}\\ 
        Family & Instrument & Train & Validation & Test & P & R & F$_1$ & P & R & F$_1$\\
        \midrule
        \midrule
         \multirow{7}{2.5cm}{Strings} & Violin & 1277 & 150 & 75 & 0.43 & 0.65 & 0.51 & 0.33 & 0.49 & 0.40\\
         & Viola & 828 & 49 & 97 & 0.78 & 0.70 & 0.74 & 0.47 & 0.38 & 0.42\\
         & Cello & 756 & 44 & 89 & 0.56 & 0.46 & 0.51 & 0.30 & 0.28 & 0.29\\
         & Double-bass & 724 & 43 & 85 & 0.79 & 0.59 & 0.68 & 0.43 & 0.52 & 0.47\\
         & Guitar & 90 & 5 & 11 & \textbf{1.00} & 0.73 & \textbf{0.84} & 1.00 & 0.18 & 0.31\\
         & Banjo & 63 & 4 & 7 & 0.33 & 0.14 & 0.20 & 0.00 & 0.00 & 0.00\\
         & Mandolin & 68 & 4 & 8 & 0.67 & 0.50 & 0.57 & 0.00 & 0.00 & 0.00\\
         \midrule
         \multirow{7}{2.5cm}{Woodwinds} & Clarinet & 719 & 42 & 85 & 0.63 & 0.56 & 0.60 & 0.50 & 0.06 & 0.11\\
         & Bass-clarinet & 802 & 48 & 94 & 0.81 & 0.49 & 0.61 & 0.75 & 0.10 & 0.17\\
         & Saxophone & 623 & 37 & 73 & 0.46 & 0.33 & 0.38 & 0.08 & 0.07 & 0.07\\
         & Flute & 746 & 44 & 88 & 0.66 & 0.76 & 0.71 & 0.43 & 0.07 & 0.12\\
         & Oboe & 507 & 29 & 60 & 0.68 & \textbf{0.87} & 0.76 & 0.45 & 0.82 & 0.58\\
         & Bassoon & 612 & 36 & 72 & 0.88 & 0.53 & 0.66 & 0.66 & 0.49 & 0.56\\
         & Contrabassoon & 604 & 35 & 71 & 0.90 & 0.51 & 0.65 & 0.46 & 0.37 & 0.41\\
         & English-horn & 587 & 35 & 69 & 0.69 & 0.59 & 0.64 & 0.47 & 0.72 & 0.57\\
         \midrule
         \multirow{5}{2.5cm}{Brass} & French-horn & 554 & 33 & 65 & 0.62 & 0.69 & 0.65 & 0.52 & 0.66 & 0.59\\
         & Trombone & 706 & 42 & 83 & 0.35 & 0.64 & 0.45 & 0.29 & 0.61 & 0.39\\
         & Trumpet & 412 & 25 & 48 & 0.82 & 0.77 & 0.80 & 0.60 & 0.73 & 0.66\\
         & Tuba & 826 & 49 & 97 & 0.76 & 0.81 & 0.79 & 0.56 & 0.67 & 0.61\\
         \midrule
         \multicolumn{2}{c|}{Chromatic Percussion} & 128 & 15 & 5 & 0.17 & 0.40 & 0.24 & 0.09 & 0.20 & 0.13\\
         \midrule
         \midrule
         \multicolumn{2}{c|}{Total} & 11630 & 694 & 1357 & \textbf{0.66} & \textbf{0.62} & \textbf{0.62} & 0.45 & 0.41 & 0.38\\
        \bottomrule
         %\cmidrule{3-4}
    \end{tabular}
    \caption{Number of instruments per class and classification metrics of each instrument. Best results are highlighted.}
    \label{tab:results}
\end{table*}

For our best-performing models, with $h$=8, we also show the confusion matrices in Fig. \ref{fig:confusion_matrices} along with the metrics for each instrument in Table \ref{tab:results}. We can see that the instruments with a higher F$_1$ values are the viola with a F$_1$ value of 0.74, the oboe with F$_1$=0.76, the trumpet with F$_1$=0.80, the tuba with F$_1$=0.79 and the guitar with F$_1$=0.84.
The model confuses the instruments of the bowed strings with the woodwinds, specially when bowed strings instruments play piano or pianissimo when playing high pitches. However, bowed strings instruments such as the violin and the viola are well distinguished by the model in spite of having a similar timbre. There are only 4 out of 150 violin samples of our test set classified as violas. The family that our model find the hardest to classify is the chromatic percussion due to the variety of instruments that belong to this class and the small number of samples of these instruments in the dataset. Data augmentation of these instruments could be done to increase the results of this work. 

In Fig. \ref{fig:att_weights} we show an example of the attention weights of the att1 layer (see Table \ref{tab:models}) in the Freq.\ Attention model with $h$=8. We show 2 samples, the first one (top) is a violin, \texttt{violin\_G6\_1\_fortissimo\_arco-normal}, and the second one (bottom) a viola, \texttt{viola\_G6\_1\_fortissimo\_arco-normal}. Both instruments are taken from our validation set and they are playing the same note with the same dynamic and technique. This allows us to know how attention learns from the input magnitude spectrogram. Analyzing the weights, we can see that the attention layer focus on the frequency bins where the energy is higher, allowing the model learn from the formants in the spectrum.

%----------5.DISCUSSION-----------
\section{Discussion}\label{sec:discussion}

We have proposed a supervised model based on a multi-head attention layer which learns timbre representations from monophonic music recordings. We show that the model can distinguish between different timbres for a wide-variety of instruments, some of them with the same pitch range.

Previous works use datasets with a mix of classical and electronic instruments but not with all the instruments that are in a classical music orchestra which present more similar timbres and which we address with the dataset we use in our work. The dataset we use in this work not only presents more sound samples than other datasets but it has also more variety of playing techniques and dynamics.

Analyzing the results of this work, we can affirm that the attention mechanism \cite{attention} not only improves the results of other neural network architectures such as fully connected layers, but its number of parameters is lower than fully connected or Long-Short Term Memory (LSTM) architectures. We can also affirm from the results that adding attention heads to the attention layer up to a certain limit (8 heads) improves the performance of the model. 
Analyzing the weights and the activation maps for different inputs in Fig. \ref{fig:att_weights}, we can see that the model learns timbre and that it distinguish between different instruments of the same family (bowed strings in Fig. \ref{fig:att_weights}) which play the same note (same pitch).
The results in Table \ref{tab:results} and the confusion matrices in Fig. \ref{fig:confusion_matrices} show that the model confuses instruments which timbre is different because they are instruments from different instrument families, but their pitch range is similar. An example of that is the the english horn (woodwind) or the cello (bowed strings) which are confused with the trombone (brass). However, other instruments that belong to the same instrument family, thus, instruments with similiar timbres such as the violin and viola are well identify by the model.
In spite of working with a monophonic samples of instruments with similar timbres and not with mixed signals composed by very different timbres as recent research does, our work reaches F$_1$ values of previous works in determined instruments such as the clarinet, and outperforms trumpet and flute \cite{GururaniSL19}. Where our model does not reach previous works results is in instruments like the cello or the guitar, due to the fact that we train our model with a dataset of very similar timbres.

If we compare our model to classical methods which use the same number of instruments, we see that the accuracy obtained by Agostini et al. \cite{AgostiniLP03} was 78.6\% for 20 instrument classes using a SVM with Radial Basis Functions (RBF) kernel versus the 62\% accuracy of our end-to-end method. The accuracy of our method does not reach classical methods results due to the fact of the variety of playing techniques that are present in our samples. Agostini et al. used pizzicato and bowed techniques while we are using spiccato, martele or glissando besides those techniques. Our work is also an end-to-end method which does not need the pre-processing steps proposed by Agostini et al.

%----------6.CONCLUSIONS-----------
\section{Conclusions and Future Work}\label{sec:conclusions}

Learning timbre with end-to-end models is an open research area in MIR. This work could help future research to learn timbre of instruments of the same families which play notes in the same pitch range.
We also give proofs of the difficulties that the model has with some instruments of the orchestra due to the unbalanced samples in the dataset, so future research should perform data augmentation on those instruments.
With this work we show that attention not only helps deep learning models to better understand timbre but it also requires less parameters. However, there are still some timbres that this model do not distinguish. Future work should focus on building architectures that combine attention with layers that help the model learn temporal spectral features and also use these architectures for unsupervised learning to better disentangle pitch and timbre. Performing small changes to our model and training it with other loss functions as Gururani et al. proposed \cite{GururaniSL19}, our work could also be used in polyphonic recordings to extract the most predominant instruments.

%----------7.ACKNOWLEDGEMENTS-----------
%\import{sections/}{7-acknowledge.tex}

% For bibtex users:
\bibliography{main}

\end{document}